\documentclass[journal]{IEEEtran}

\normalsize

\ifCLASSINFOpdf
\else
\fi

\usepackage{verbatim} 
\usepackage[dvips]{graphicx} 
\usepackage[cmex10]{amsmath}
\usepackage{amssymb,amsfonts}
\usepackage{mathtools}  
\usepackage{framed}
\usepackage{lipsum}
\usepackage{mdframed}
\usepackage{adjustbox}

\usepackage{cases}
\usepackage{theorem} 
\def\myproof{\noindent{{\textbf{Proof:}}}} 

\ifdefined\myproof
\else
\def\myproof{\proof}

\fi

\newtheorem{proposition}{Proposition}

\newtheorem{remark}{Remark}

\newtheorem{proof}{Proof} 

\usepackage{cite} 

\usepackage[tight,footnotesize]{subfigure}

\usepackage[linesnumbered,ruled]{algorithm2e}
\usepackage{multirow}

\begin{document}

%
\title{  Antenna Optimization for Decode-and-Forward Relay in Magnetic Induction Communications \vspace{-0.1em}}
%
%
%
%

\author{Honglei~Ma,
        Erwu~Liu, ~\IEEEmembership{Senior Member,~IEEE,}
        Rui~Wang, ~\IEEEmembership{Senior Member,~IEEE,}
        Xuefeng~Yin, ~\IEEEmembership{Member,~IEEE,}
        Zhibing~Xu,
        Xinyu~Qu,
        and~Bofeng~Li \vspace{-0.0em}
\thanks{This is the author?s accepted manuscript, which has been published in IEEE Transactions on Vehicular Technology. The final published version can be accessed at https://doi.org/10.1109/TVT.2019.2963357. Copyright of the published work belongs to IEEE.}
\thanks{This work is supported by National Science Foundation China under Grants (61571330 and 61771345)(Corresponding author: Erwu~Liu).}
\thanks{Honglei~Ma, Erwu~Liu, Rui~Wang, Xuefeng~Yin, Xinyu~Qu are with the School of
Electronics and Information Engineering, Tongji University, Shanghai 201804, China, E-mail: holyma@yeah.net, erwu.liu@ieee.org, ruiwang@tongji.edu.cn,  549197567@qq.com.  }
\thanks{Zhibing~Xu is with Huawei Technologies Co., Ltd., E-mail: marco.xuzhibing@huawei.com.}
\thanks{Bofeng~Li is with the College of Surveying and Geo-Informatics, Tongji University, Shanghai 200092, China, bofeng\_li@tongji.edu.cn.}
\thanks{Digital Object Identifier 10.1109/TVT.2019.2963357}
}

%
%

\markboth{}%
{Submitted paper}
%



\maketitle

\begin{abstract}
Magnetic Induction (MI) communication is effective in underground tunnels for emergency rescue vehicle due to the small-size antenna. It can highly benefit from a cooperative decode-and-forward (DF) relay to achieve a higher data rate. However, its channel gain is extremely position-and-orientation-selective. The unreachable space increases the  complexity of the antenna deployment.  To find the best antenna position and orientation (PO) of the relay achieving the higher data rate,  this paper formulates the optimization problem of the relay MI antenna PO  with  tunnel constraints. To solve the problem more quickly, we propose to use geometric modeling to eliminate the tunnel constraints and develop a random-search algorithm achieving a fast convergence and excellent global search ability. Simulations show that the proposed algorithm can quickly converge to one optimum which signifies a noticeable improvement of data rate for vehicle MI systems with weak signals.
\end{abstract}


%
\IEEEpeerreviewmaketitle

\vspace{-0.0em}
\section{Introduction}\label{sect_intro}
\vspace{-0.0em}
%
%
%
%

When  disaster strikes, the communication system in underground tunnels may be damaged. Therefore, a simple through-the-earth communication device is necessary for an underground rescue vehicle to communicate with each other. The  magnetic induction (MI) communication  has been proven to be a useful for a wide variety of underground applications \cite{Kisseleff2018Survey}. For the through-the-earth communications, there are mainly two methods for a long-range transmission. One method is the MI waveguide which  extends the transmission range by deploying  additional passive relays between two transceiver nodes\cite{Sharma2017Magnetic }. However, this method may not be  suitable for  emergency applications since it may take a lot of rescue time for deploying  a large number of  passive relays within the tunnels. The other method is the decrease in signal frequency $f$ since the low frequency (1--10KHz) can achieve small skin depth $\delta\approx\sqrt{\frac{1}{\pi f\mu \sigma}}$\cite{kisseleff2013channel} and reduce the eddy-current loss in the long-distance underground material. In 2012,  researchers  developed an MI device  achieving an underground transmitting distance of 310m\cite{zhang2014cooperative}.  However, the transmission rate of the MI communication is extremely low due to the narrow bandwidth.

 Recently, a few studies have proven that the use of an active MI relay setting up a cooperative MI system  can boost overall data rate up to an order of magnitude\cite{Kisseleff2018Survey}\cite{Kisseleff2015On}.  With advantage over electromagnetic wave channels, MI channels have no small-scale fading and multipath effect.  In\cite{Kisseleff2015On}, researchers investigate the capacity improvement by using an active relay with amplify-and-forward and DF  cooperative schemes achieving $10^3$\% data rate increase. Under the DF cooperative scheme, the relay decodes the received signals and then re-codes and transmits them to the destination\cite{Li2019}.  In \cite{Li2019}, researchers  propose the hybrid structure that combines the waveguide and active relaying transmission techniques. However, they also need to deploy many passive relays.

\begin{figure}[b]
        \centering
        \vspace{-1.6em}
        \includegraphics[width=2.1in, height=1.5in]{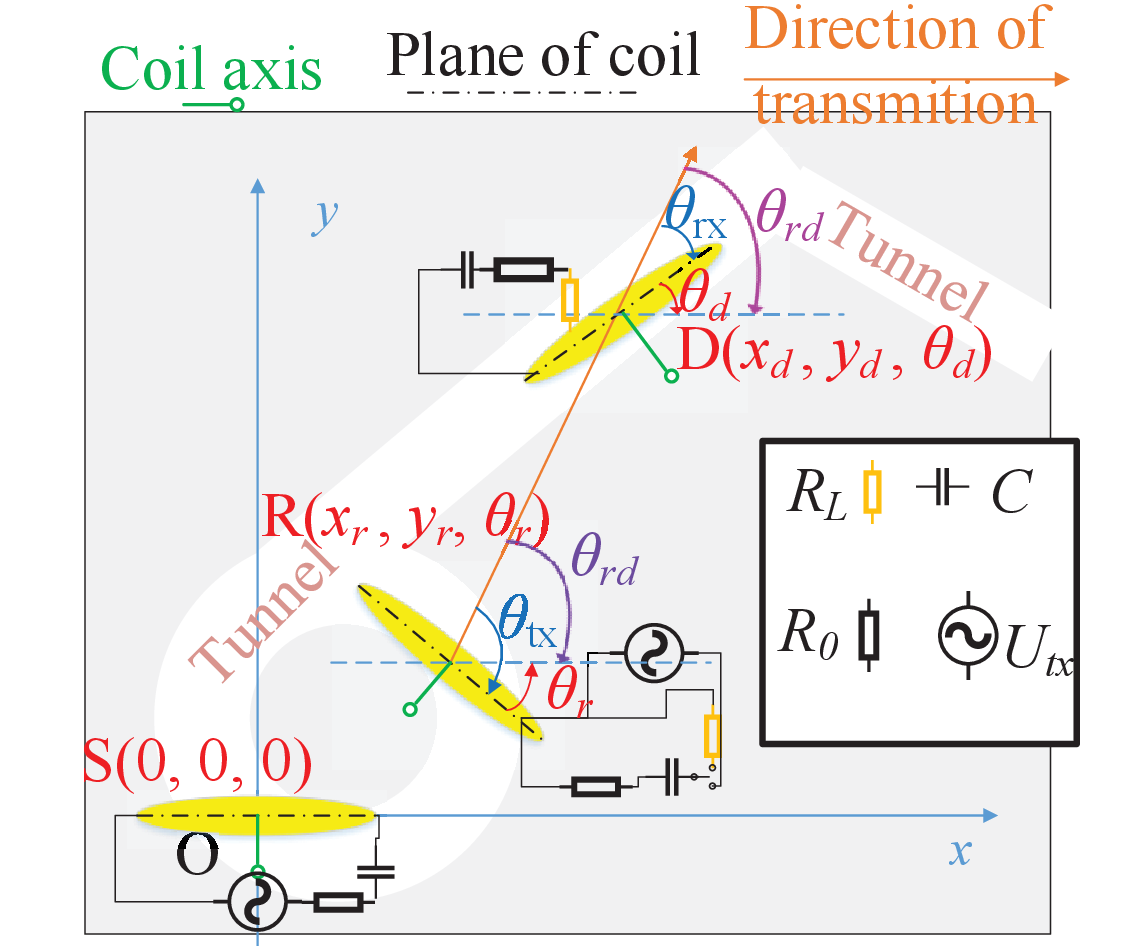}\\
        \caption{A cooperative MI network with a single relay.
 }\label{fig_miAngle}
 \end{figure}

 The mutual inductance between arbitrarily placed coils R and D (see Fig.\ref{fig_miAngle}) can be given by\cite{Tan2015Environment}
 \vspace{-0.5em}
   \begin{equation}\label{eqn_mrd0}
\begin{aligned}
 M_{rd} =\mu\pi n^2 \frac{a^4 \mathcal{G}_{e}}{2r_{rd}^3}\left(\sin \theta_\text{rx} \sin \theta_\text{tx} - \frac{1}{2} \cos \theta_\text{rx} \cos \theta_\text{tx}\right),
 \end{aligned}
   \vspace{-0.5em}
 \end{equation}
where $\mu$ is the permeability of medium, $\theta_\text{tx}$  and $\theta_\text{rx}$ are the angles between the coil radial directions and the line connecting the two coil centers, respectively, $r_{rd}$ is the distance between two coils,  $\mathcal{G}_e$=exp$(\frac{-r_{rd}}{\delta})$ is an additional loss factor due to eddy current (skin effect), $\sigma$ is the conductivity of medium.  Since $\theta_\text{tx}$ and  $\theta_\text{rx}$ depend on the antenna POs of R and D, the antenna PO  has a significant effect on the system capacity. Therefore, besides the transmission power and frequency optimizations\cite{Kisseleff2015On}, the antenna PO optimization for the active relay is also important. However, the work\cite{Kisseleff2015On} does not theoretically analyze  the antenna PO optimization.

Until now, there are very limited works theoretically analyzing the antenna PO optimization on the active MI relay. For the position optimization, the unreachable spaces need to be considered since it is impossible to deploy the relay in these spaces. In \cite{zhang2014cooperative}, as researchers only optimize the antenna orientation of a cooperative MI node at a special position, they do not need to consider the unreachable spaces. Hence, their optimization problem is unconstrained.
Moreover, they do not model the data forwarding process since they aim at achieving a potential communication range instead of data rate. Thus, the objective function of their optimization problem is simplified as a simple cosine function.

This paper studies the  antenna PO optimization problem for an active MI relay in a cooperative MI system. The difference between our network topology and that of \cite{Kisseleff2015On} is that the source and destination in our topology can communicate with each other. We adopt the DF scheme as the data forwarding scheme to achieve a better data rate\cite{Kisseleff2015On}.  Also, different from \cite{zhang2014cooperative}, this paper focuses on the data rate improvement brought by an unfixed-position relay. We should model the data forwarding process. The optimization problem is high-complexity and non-convex. More importantly,  this paper considers that the relay  should be deployed  within tunnels modeled as lines which are expressed by equations. Thus, the optimization problem is with equality constraints called \emph{tunnel constraints} which may sharply increase the time complexity of the problem solving.

 Our contributions are as follows: 1) We formulate the relay antenna PO optimization problem  with a tunnel constraint to obtain the best antenna PO of the DF relay for potentially increasing the data rate of the MI system.  2) We propose a new  method to eliminate the constraint from this optimization problem. After eliminating the constraint,  the time complexity sharply decreases.  3) We propose a novel algorithm to solve this optimization problem. Such algorithm achieves a faster convergence and better global search ability than widely used optimization algorithm. Thus, the time of the antenna deploying would further decrease. Simulation results show that the algorithm can quickly converge to one optimum which can achieve a remarkable increase in data rate.

\vspace{-0.8em}
\section{System Model and Problem Formulation}\label{sect model}
 \vspace{-0.2em}
As shown in Fig.\ref{fig_miAngle},  we assume a working MI link (S$\rightarrow $D) deployed in a tunnel. For the long-range  communication, the diameter of the tunnel is much smaller than the underground environment. Thus, we model such environment as an infinite plane with homogeneous materials and establish a Cartesian coordinate system with the origin S for the plane.  The tunnel can be treated as a line expressed by an equation $U(x, y)$$=$$0$ called \emph{tunnel line}.  Node D is the destination with the specific antenna PO $\mathbf{v}_d$$=$$(x_d, y_d, \theta_d)$. Hereafter, the antenna PO of node $(\cdot)$ is expressed as a triple $(x_{(\cdot)}, y_{(\cdot)}, \theta_{(\cdot)})$,  where $(x_{(\cdot)}, y_{(\cdot)})$ is the antenna position, $\theta_{(\cdot)}$ is the antenna orientation which is the angle between the x axis and the radial direction. Node S is the source with the antenna PO $\mathbf{v}_s$$=$$(0, 0, 0)$. We deploy a DF relay to establish a cooperative MI system (DF system) with S and D for data rate increasing. The antenna PO of the relay is $\mathbf{v}_r$$=$$(x_r, y_r, \theta_r)$  to be optimized. The relay should be deployed within the tunnel.  Similar to\cite{Kisseleff2015On},  every node coil (antenna) with resistance $R_0$ has a radius of $a$ and the number of turns $n$. The capacitor with capacitance $C$ is tuned to make the circuit resonance at our target signal frequency  $f_0$ =$1/2\pi\sqrt{{L_0}C}$, where ${L_0}$ is the inductivity of the coil. $R_L$ is the load resistor for minimum power reflections\cite{Kisseleff2015On}.

\emph{Conventions}: In this paper, the antenna PO notations $(\mathbf{v}_r)$ and $(v_r)$ are equivalent to the tuples $(x_r, y_r, \theta_r)$ and $(x_r, \theta_r)$, respectively.  The subscripts  `$s$', `$r$' and `$d$' represent the \emph{source}, \emph{relay} and \emph{destination}, respectively.  

\vspace{-0.3em}
\subsection{Mutual Inductance and Antenna PO}\label{subsect_model_optm}
\vspace{-0.3em}
The mutual inductance between R and D is given by~\eqref{eqn_mrd0}.  Fig.\ref{fig_miAngle} shows that $\theta_{\textmd{rx}}$$=$$\theta_{rd}$$-$$\theta_d$ and   $\theta_{\textmd{tx}}$$=$$\theta_{rd}$$-$$\theta_r$. According to the  fundamental analytic geometry, we obtain $\theta_{rd}$$=$$\tan^{-1}(\tfrac{y_r-y_d}{x_r-x_d})$, $\sin \theta_\text{rx} \sin \theta_\text{tx}$$=$$\frac{1}{2}(\cos (\theta_{\mathrm{rx}}-\theta_{\mathrm{tx}})-\cos (\theta_{\mathrm{rx}}+\theta_{\mathrm{tx}}))$, $\cos \theta_{\mathrm{rx}} \cos \theta_{\mathrm{tx}}$$=$$\frac{1}{2}(\cos (\theta_{\mathrm{rx}}+\theta_{\mathrm{tx}})+\cos (\theta_{\mathrm{rx}}-\theta_{\mathrm{tx}}))$. We substitute these equations into ~\eqref{eqn_mrd0} and get
\vspace{-0.4em}
\begin{equation}\label{eqn_mrd}
\begin{aligned}
&M_{rd}(\mathbf{v}_r, \mathbf{v}_d) =   \\
  & \frac{\kappa\left[ \cos( \theta_d-\theta_r)-3\cos\left(\theta_d+\theta_r-2\tan^{-1}(\frac{y_r-y_d}{x_r-x_d})\right) \right]}
   {e^{\sqrt{(x_r-x_d)^2 + (y_r-y_d)^2}/\delta} 8\sqrt{\left[(x_r-x_d)^2+(y_r-y_d)^2\right]^3}},
 \end{aligned}
 \end{equation}
 where $\kappa$  = $\mu \pi n^2 a^4$. When we replace $\mathbf{v}_r$ and $\mathbf{v}_d$   with $\mathbf{v}_s$ and $\mathbf{v}_r$ in~\eqref{eqn_mrd}, we get the mutual inductance between S and R:
  \vspace{-0.4em}
\begin{equation}\label{eqn_msr}
\begin{aligned}
 &M_{sr}(\mathbf{v}_r) = \frac{\kappa\left[ \cos( \theta_r)-3\cos\left(\theta_r-2\tan^{-1}(\frac{y_r}{x_r})\right) \right]}
   {e^{\sqrt{x_r^2 + y_r^2}/\delta} 8\sqrt{\left(x_r^2+y_r^2\right)^3}} \\
    &=  \frac{\kappa\left[ cos\theta_r-3\left(\frac{1-y_r^2/x_r^2}{1+y_r^2/x_r^2}\cos\theta_r + \frac{2y_r/x_r}{1+y_r^2/x_r^2}\sin\theta_r\right) \right]}
   {e^{\sqrt{x_r^2 + y_r^2}/\delta} 8\sqrt{\left(x_r^2+y_r^2\right)^3}} \\
   &=\!\kappa \frac{\left( 2y_r^2-x_r^2\right) \cos \theta_r\!-\!3x_r y_r\sin \theta_r }
   {4e^{\sqrt{x_r^2\!+\!y_r^2}/\delta} \sqrt{\left(x_r^2+y_r^2\right)^5}}.
 \end{aligned}
 \vspace{-0.4em}
 \end{equation}
 Accordingly, the mutual inductance between S and D is $M_{sd} = M_{sr}(\mathbf{v}_d)$. For the specific destination, its antenna PO $\mathbf{v}_d$$=$$(x_d, y_d, \theta_d$) is a constant vector  and $M_{sd}$ is a constant.

\vspace{-0.0em}

\subsection{Antenna PO Optimization Formulation}\label{subsect_APO}
\vspace{-0.2em}
To obtain the optimal antenna PO, the expression of cooperative MI channel power gain with respect to antenna PO is needed. In  through-the-earth scenarios, due to the sufficiently long transmission distance and low signal frequency, the current induced by other node coils can be ignored, and the DF relay based network model can be simplified as the wireless ad-hoc network model similar to\cite{lin2015distributed}.  As shown in Fig.\ref{fig_miAngle} and given the signal frequency $f$, the impendence of the transmitter and receiver circuits are $Z_\text{tx}$ = $2\pi f {L_0} + \frac{j}{2\pi f C} + R_0 +R_L$ and $Z_\text{rx}$ = $Z_\text{tx}$, respectively. According to \cite{lin2015distributed} and let $h_0(f) =|\frac{(2\pi f)^2 R_L}{ Z_\text{rx}^2 Z_\text{tx}}|$, the channel power gains between all node pairs are $H_{sd}=h(f) M_{sd}^2$,  $H_{sr}(\mathbf{v}_r)=h(f) M_{sr}^2(\mathbf{v}_r)$ and $H_{rd}(\mathbf{v}_r)=h(f) M_{rd}^2(\mathbf{v}_r)$.

The expression of achievable rate of DF system  without diversity combining (combining the multiple independent received signals  to decode the message) is given in\cite{Kisseleff2015On}. However, due to the existence of direct link (S $\rightarrow$ D), we should consider the diversity combining.  As depicted in\cite{Hong2010Cooperative}, the diversity is achieved via two phases. In phase I, S broadcasts its signal to D and R, and the relay R decodes the received signal with the data rate $C_{\rm I}$.  In phase II, R forwards the decoded signal to D. By using the maximal ratio combining (MRC), the equivalent data rate in phase II is $C_{\rm II}$. Based on\cite{Hong2010Cooperative}, we derive the expression of achievable rate of DF based cooperative system $Q_{\rm DF}$ as
\begin{subequations}\label{eqn_Cdf}
\vspace{-0.5em}
\begin{align}
   Q_{\rm DF}(\mathbf{v}_r) &\!=\! \left[\iota_{\rm DF}(\mathbf{v}_r)C_{\rm DF} (\mathbf{v}_r)\! -\!\delta_{\rm DF}\right]^+,  \tag{4a} \label{eqn_cdf:qdf}\\ %
   C_{\rm DF} (\mathbf{v}_r) &= \frac{\Gamma_{\rm fec}}{2}\min\{C_{\rm I}(\mathbf{v}_r), C_{\rm II}(\mathbf{v}_r)\}, \tag{4b}  \label{eqn_cdf:cdf}
\end{align}
\end{subequations}

\begin{subequations}\label{eqn_Cdf1}
\vspace{-0.5em}
\begin{align}
   C_{\rm I}(\mathbf{v}_r)&= C_{sr}(\mathbf{v}_r)       =\int_B\log_2\left(1 + P_{s}\frac{H_{sr}(f, \mathbf{v}_r)}{P_N}\right)df, \notag \\
     &=  \int_B\log_2 \left(1 +  P_s\frac{h(f)M^2_{sr}(\mathbf{v}_r)}{P_N}\right)df, \tag{4c} \label{eqn_csr}\\    %
   C_{\rm II}(\mathbf{v}_r)\!&= \!\int_B\log_2  \left(1\!+\!h(f)\tfrac{P_rM^2_{rd}( \mathbf{v}_r)\!+\!P_sM^2_{sd} }{P_N}\right)df, \tag{4d} \label{eqn_csrd}%
\end{align}
\end{subequations}

\setcounter{equation}{4}
where $[\cdot]^+$ denotes $\max\{\cdot, 0\}$, $P_s$ and $P_r$ are the transmit powers at S and R, respectively, $P_N$ is the ambient noise power,  $B$ is the 3-dB bandwidth, the constant $\Gamma_{\rm fec}$ is the gap between the actual data rate and Shannon's capacity,  the constant $\delta_{(\cdot)}$  is the data rate loss brought by the time delay due to the cross-layer protocol\cite{lin2015distributed}, $\iota_{\rm DF}(\mathbf{v}_r) = 1 - p^{(e)}_{\rm DF}$ where $p^{(e)}_{(\cdot)}$ denotes the bit error ratio. For a comparison with the direct MI transmission, we assume that the total transmit power  of the DF system equals to that of the direct MI system, \emph{i.e.}, $P_{\rm tot} = (P_s + \frac{P_r}{\Gamma_p})$ where $\Gamma_p$ is a constant  to compensate the gap between the receive power and processing power of the relay.  According to\cite{lin2015distributed}, we get the achievable rate of direct MI system as
 \begin{equation}\label{eqn_Csd}
\begin{aligned}
Q_{sd} \!= \!\left[\iota_{sd} \Gamma_{\rm fec}  \int_B\log_2 \left(1\!+\!\tfrac{P_{\rm tot}}{P_N}h(f)M^2_{sd}\right)df
       -\delta_{sd}\right]^+,
\end{aligned}
\end{equation}
where $\iota_{sd} = 1 - p^{(e)}_{sd}$.  $Q_{sd}$ is a constant as $M_{sd}$ is a constant.

Next, we formulate the optimization problem of the relay antenna PO. Since  S and  D  can communicate with each other directly, we should guarantee $p^{(e)}_{sd}$$\ll$$1$. Suppose R$^*$ is the optimal relay with the antenna PO $\mathbf{v}_r^*$$=$$(x_r^*, y_r^*, \theta^*_r)$.  R$^*$ can also communicate with S and D, \emph{i.e.,} $p^{(e)}_{sr}$$\ll$$1$ and $p^{(e)}_{rd}$$\ll$$1$. Hence,  $\iota_{\rm DF}^*$ is a constant for $p^{(e)}_{\rm DF}$$\simeq$$ p^{(e)}_{sr}$$+$$ p^{(e)}_{rd}$$-$$2 p^{(e)}_{sr} p^{(e)}_{rd}$\cite{Wu2011BER}. Moreover, since $C_{\rm DF}(x^*_r, y^*_r, \theta^*_r)$$\geq$$C_{\rm DF}(x_r, y_r, \theta_r)$ is always satisfied, we can define a function $G(\mathbf{v}_r)$$=$$G(x_r, y_r, \theta_r)$ (called \emph{cooperative quasi-gain} function) as the optimization goal for the relay antenna:
\vspace{-0.1em}
 \begin{equation}\label{eqn_G}
\begin{aligned}
       G(x_r, y_r, \theta_r) &= \frac { \left[\iota^*_{\rm DF}C_{\rm DF} (x_r, y_r, \theta_r)\! -\!\delta_{\rm DF}\right]^+}{ Q_{sd} }.
\end{aligned}
\end{equation}
Let $g(x_r,y_r,\theta_r) = -G(x_r, y_r, \theta_r)$, we formulate the optimization problem of the relay antenna PO as
  \begin{subequations}\label{eqn_APOO}
\begin{align}
    &\arg \min_{x_r,y_r,\theta_r}  g(x_r,y_r,\theta_r) \label{eqn_APOO:g} \\
    & s.t. \ \ \ U(x_r, y_r) = 0, \label{eqn_APOO:u}
\end{align}
\end{subequations}
where we define the constraint~\eqref{eqn_APOO:u} as a \emph{tunnel constraint}.

 \vspace{-0.8em}
\begin{remark}\label{rmk1}
The problem~\eqref{eqn_APOO} is non-convex and has a large number of local optimums near the curve $C_{sr}(\mathbf{v}_r)=C_{\rm II}(\mathbf{v}_r)$.
\end{remark}
\vspace{-0.4em}

In some tunnels with relatively large width, as the eddy-current losses might be slightly reduced due to the air conductivity, the tunnel width might have a little effect on the optimal solutions. Such an effect is related to the geometry of tunnels and the propagation of near-field signals in inhomogeneous media which will be subject to future investigations.

The MI node in most emergency applications is an embedded device with low computational abilities. Therefore, it is necessary to propose an antenna PO optimization algorithm which aims at improving the convergence performance and global search ability (chance to obtain the global optimal solution).

\vspace{-0.5em}

\section{ Antenna Position and Orientation Optimization }\label{sect APO}
In this section, to eliminate the constraint~\eqref{eqn_APOO:u}, we first propose the  low-complexity method of geometric approximation which reduces the dimension of the objective function in~\eqref{eqn_APOO:g}. Then, we propose the antenna optimization algorithm with fast convergence and excellent global search ability.
\vspace{-0.8em}
\subsection{Tunnel Constraint Elimination}\label{subsect_model_gain}
 The antenna PO optimization problem~\eqref{eqn_APOO} is non-convex and has an equality constraint. The solution of any non-convex constrained optimization problem is challenging. The general approaches to eliminate the constraints are the method of Lagrangian's multipliers and penalty functions\cite{Yu2018On}, \emph{e.g.}, the recent literature \cite{Yu2018On} provides dual subgradient/gradient algorithm as
\begin{subequations} \label{eqn_maxg_lagrangian}
    \begin{align}
        \mathbf{v}^{(k)}_{xy} &= \arg\min\bigg(g(\mathbf{v}_r^{(k)}) + \lambda^{(k)}U(\mathbf{v}^{(k)}_{xy})\bigg), \label{eqn_maxg_lagrangian:f}\\
        \lambda^{(k+1)} &= \max\left(\lambda^{(k)} + \varsigma U(\mathbf{v}^{(k)}_{xy}), 0  \right), \label{eqn_maxg_lagrangian:lda}
    \end{align}
\end{subequations}
where $k$ represents the $k$-th iteration, $\varsigma$ is the step size of $\lambda$, $(\mathbf{v}^{(k)}_{xy})$ = $(x_r^{(k)},y_r^{(k)},0)$. However, as we should both ensure the convergence of~\eqref{eqn_maxg_lagrangian:f} and~\eqref{eqn_maxg_lagrangian:lda},  the phenomenon of Remark \ref{rmk1} causes the dramatically poor convergence performance. Thus,  the method of Lagrangian's multipliers based on subgradient is not practicable for the optimization problem~\eqref{eqn_APOO}.

Therefore, we develop a low complexity approach called geometric approximation to  eliminate the constraint. If we attempt to solve~\eqref{eqn_APOO:u} for the variable $y_r$, we can obtain the function of the $x_r$, \emph{i.e.,} $y_r $$=$$ U^{-1}(x_r, 0)$$\triangleq$$ \mathcal{Y}(x_r) $ where $U^{-1}(\cdot)$ is the inverse function of  $U(\cdot)$. Then, we substitute $y_r$$=$$\mathcal{Y}(x_r)$ into~\eqref{eqn_APOO:g} and get the unconstrained optimization problem  $\arg \min_{x_r,\theta_r} g(x_r,\mathcal{Y}(x_r), \theta_r)$.

   \begin{figure}[ht]
        \centering
        \vspace{-1.0em}
        \includegraphics[width=2.0in, height=1.4in]{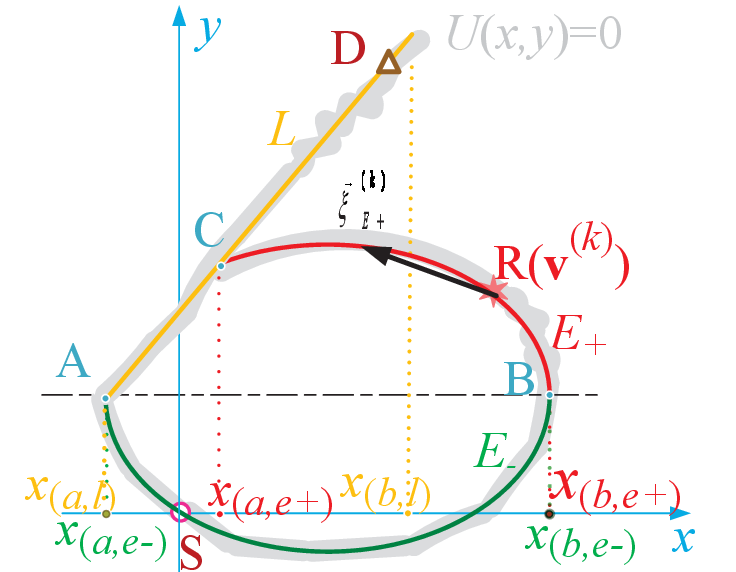}\\
        \caption{A geometric approximation of the tunnel. Sub-tunnel ACD is fitted into a straight segment $L$ with the start $(x_{a,l}, \mathcal{Y}_l(x_{a,l}))$ and the end $(x_{b,l}, \mathcal{Y}_l(x_{b,l}))$. Sub-tunnel ASBC is fitted into an ellipse arc $(E_+\bigcup E_-)$. $\vec{\xi}_{E_+}^{(k)}$ is the step vector of searching along the arc $E_+$ at the $k$-th iteration. R$(\cdot)$ is the relay being optimized (called prober).
          \vspace{-0.8em}
 }\label{fig_road}
    \end{figure}

However, it may be difficult to obtain $\mathcal{Y}(x_r)=U^{-1}(x_r, 0)$ since there may exist  multiple solutions of~\eqref{eqn_APOO:u} for $y_r$. Considering that any line with non-zero curvature and zero curvature can be fitted into an arc of an ellipse and a straight line segment, respectively, we approximately model  the tunnel line as the combination of  $n_e$ arcs of ellipses and $n_l$ straight line segments, see Fig.\ref{fig_road}.  Without loss of generality, we set $n_e=1$  and $n_l=1$ in this paper.
According to  elementary geometry, the straight line segment can be given by $L = \{(x_r, y_r)| y_r = \eta x_r + \beta, x_{a,l}< x_r < x_{b,l}\}$ where $\eta$ and $\beta$ are the slope and intercept, respectively.  After estimating the long (short) axis $R_{a}$, short (long) axis $R_{b}$ and the center ($x_{c},y_{c}$), the  ellipse can be expressed by the equation $\frac{(x_r-x_c)^2}{R_a^2} + \frac{(y_r-y_c)^2}{R_b^2} = 1$. By solving this equation for $y_r$ to get $U^{-1}(x_r, 0)$, we obtain the expressions of two arcs of the ellipse,  \emph{i.e.,} $E_+ = \{(x_r, y_r)|\ y_r= y_{c} + R_{b}(1 - \tfrac{(x_r-x_{c})^2}{R^{2}_{a} })^{\frac{1}{2}},  x_{a,e+}< x_r < x_{b,e+} \}$ and $E_- = \{(x_r, y_r)|\ y_r =  y_{c} - R_{b}(1 - \tfrac{(x_r-x_{c})^2}{R^{2}_{a} })^{\frac{1}{2}},  x_{a,e-}< x_r < x_{b,e-}\}$. The optimization problem~\eqref{eqn_APOO} is deduced as
  \begin{equation}\label{eqn_APOOW1}
\begin{aligned}
  \arg \min_{x_r,y_r,\theta_r} g(x_r,y_r, \theta_r)   \ \
  s.t. \ \ E_+ \bigcup E_-\bigcup L,
\end{aligned}
\end{equation}
which is equivalent to the union of three sub-problems:

 \vspace{-1.4em}
\begin{subequations} \label{eqn_APOUS}
  \begin{align}
    &\mathcal{P}_{i}: \   \min_{x_r,\theta_r}g_{i}(x_r, \theta_r) = \min_{x_r,\theta_r}g (x_r,\mathcal{Y}_i (x_r),\theta_r) \ \label{eqn_APOUS:subprob0} \\
    & \ \ \ \  s.t.   \ \   x_{(a,i)} \leq x_r \leq x_{(b,i)}, \  \label{eqn_APOUS:subprob1} \\
                       &\ \ \ \ \mathcal{Y}_i(x_r) = \begin{cases}
                              y_{c} + R_{b}\sqrt{1 - \frac{(x_r-x_{c})}{R^{2}_{a} }^2}  &i = E_{+},   \\
                               y_{c} -  R_{b}\sqrt{1 - \frac{(x_r-x_{c})}{R^{2}_{a} }^2}  &i= E_{-},    \\
                              \eta x_r + \beta   &i = L, \label{eqn_APOUS:con}  \\
                          \end{cases}       \\
     & \arg \min_{x_r,\theta_r, i}\{\text{sub-problem} \  \mathcal{P}_{i} \}.\label{eqn_APOUS:prob}
  \end{align}
\end{subequations}
 Since the antenna POs violating~\eqref{eqn_APOUS:subprob1} can be easily projected back to the nearest boundary, sub-problem  $\mathcal{P}_i$ becomes an unconstrained problem.  Also, in~\eqref{eqn_APOUS}, we have reduced the dimension (2 variables) compared to the constrained optimization problem~\eqref{eqn_APOO} (with 3 variables). The time of solving the antenna optimization may significantly decrease.
\vspace{-1em}
\subsection{Antenna Optimization Algorithm}\label{subsect APOUS}
The widely used method to solve the unconstrained problem is descent method  which is described by\cite{boyd2004convex}
\begin{equation} \label{eqn_descent}
  \begin{aligned}
    v_r^{(k+1)} = v_r^{(k)} + \vec{\xi}_i^{(k)} \ \ \textbf{if} \ \ g_{i}(v_r^{(k+1)})< g_{i}(v_r^{({k})}),
  \end{aligned}
\end{equation}
where $\vec{\xi}_i^{(k)}$= $t_1 \Delta {v_r}^{(k)}$ is the searching vector along the sub-tunnel $i$, $k$ represents the $k$-th iteration step, $t_1$ is the step size, $v_r^{(k)}=(x_r^{(k)}, \theta_r^{(k)})$. When $-\Delta {v_r}^{(k)}$ =$\frac{\nabla g_i(v_r^{(k)})}{||\nabla g_i(v_r^{(k)})||}\triangleq \Delta^\nabla_i {v_r}^{(k)} $, it becomes a subgradient descent (GD) algorithm. According to Remark \ref{rmk1}, the GD algorithm is highly likely to converge to local optimums. This phenomenon exists for two reasons. Firstly, the overall step size $||\vec{\xi}_i^{(k)}||$ should be small sufficiently to make  $\vec{\xi}_i^{(k)}$ close to the actual subgradient at every point between the point $v_r^{(k+1)}$ and the point $v_r^{(k)}$. Also the small $||\vec{\xi}_i^{(k)}||$ brings a slow convergence speed. Secondly, under small step size, the GD algorithm searches the optimum only through exploiting the determined direction (called \emph{exploitation})  while ignores exploring an unknown direction with certain probability called \emph{exploration}.

Aiming at overcoming the first challenge, we apply a large $||\vec{\xi}_i^{(k)}||$  at point $v_r^{(k)}$ for searching the optimum. Under the large $||\vec{\xi}_i^{(k)}||$,  the subgradient at every point between points $v_r^{(k+1)}$ and $v_r^{(k)}$ is not the  same value. Thus, we add the component $\Delta^\diamond_i v_r^{(k)}= \frac{v_r^*\!-\!v_r^{(k)}}{||v_r^*\!-\! v_r^{(k)}||}$  to correct the search direction from negative subgradient to the direction of the global optimum. Here $v_r^*$  is the estimated global optimum with a minimal negative quasi-gain $\widehat{g_i}^*(v_r^*)$. Aiming at overcoming the second challenge, we multiply the determined searching vector by  a random vector for exploration of new optimums. Moreover, to ensure the convergence,  we use a small $||\vec{\xi}_i^{(k)}||$ when the prober is close to $\widehat{g_i}^*(v_r^*)$. In this case,  we only apply the exploration as there might be many local optimums near the global optimum according to Remark \ref{rmk1}. To further improve the global search ability and estimate  $\widehat{g_i}^*(v_r^*)$,  we randomly initiate $\mathcal{M} = 20 $ probers in the search space~\eqref{eqn_APOUS:subprob1}. Then we design the update rule of each prober as~\eqref{eqn_descent}, and the searching vector $\vec{\xi}_i^{(k)}$ for each prober is designed as

\vspace{-0.9em}
\begin{subequations}\label{eqn_stepvec}
\begin{numcases}{\hspace{-6mm}\vec{\xi}^{(k)}_{i}\hspace{-1mm}\!=\!\hspace{-1mm}}
 \hspace{-1mm}\tfrac{\min\{k, \lambda_a\}}{k}  (\chi \hbar_{ba},90^\circ\vartheta  ) \ \   \tfrac{\widehat{g_i}^*(v_r^*)- g_{i}(v_r^{(k)}) }{\widehat{g_i}^*(v_r^*)}<\varepsilon_a, \label{eqn_stepvec:small}\\
 \hspace{-2mm}(\chi  \hbar_{ba}, 90^\circ\vartheta)\!\otimes\!\left(\varpi\Delta^\nabla_i v_r^{(k)}\!+\!\Delta^\diamond_i v_r^{(k)}\right)  \text{otherwise},  \label{eqn_stepvec:big}
\end{numcases}
\begin{numcases}{\hspace{-4mm}\hbar_{ba} =  x_{(b,i)} - x_{(a,i)}  \ \ \ i\in }
 \hspace{-2mm}  E_{+}, E_{-},L  \},  \label{eqn_stepvec:delta}
\end{numcases}
\end{subequations}
where  $\chi$ and $\vartheta$ are random variables between $-0.5$ and $0.5$,  the constant $|\varepsilon_a|$$\leq$$1$, $\lambda_a$ is the minimal iterations guaranteeing the searching vector with large step size, $\varpi$ is the weight of  subgradient direction, $\otimes$ denotes element-wise vector multiplication,   the random vector $(\chi \hbar_{ba},90^\circ\vartheta)$ represents the exploration.   Here $(\hbar_{ba}, 90^\circ)$ is to sufficiently amplify the step size. The vector $\varpi\Delta^\nabla_i v_r^{(k)}$$+$$\Delta^\diamond_i v_r^{(k)}$ represents the exploitation. In general, the  searching vector is with small step size in~\eqref{eqn_stepvec:small} and with large step size in~\eqref{eqn_stepvec:big}. Finally, we propose the relay antenna PO optimization algorithm as depicted in Algorithm \ref{alg_apows}. Next, we focus on the convergence of the algorithm.

\vspace{-0.8em}
\begin{algorithm}[h]\label{alg_apows}
 \KwIn{ $\varepsilon_a$ is the threshold for the large step vector.}
  \tcc{ $v_r[m][k]=(x_r^{(k)},\theta_r^{(k)}$) for $m$-th prober}
\While{$k$++ $<$ $MaximalIterations$}
{
       \For{$m=1$ to $\mathcal{M}$ do}{
           $\widehat{g_i}^*(v_r^*)$ = $\min\limits_{m' < \mathcal{M}} g_{i}(v_r[m'][k])$ \\
           obtain $\vec{\xi}^{(k)}_{i}$ according to~\eqref{eqn_stepvec};\\
           update $v_r[m][k + 1]$ according to~\eqref{eqn_descent}; \\
       }
}
\Return the optimal solutions $\widehat{g_i}^* $ and ${v_r^*}$;
\caption{Antenna optimization algorithm. }
\end{algorithm}
\vspace{-2.0em}
\begin{proposition}\label{prop_ADOA}
 Regardless of any $\mathcal{M}$ initial APOs chosen, Algorithm \ref{alg_apows} converges.
\end{proposition}
\vspace{-0.5em}
\begin{myproof}\label{prfAdo}
Since $g_{i}(v_r^{(k+1)})\!< \!g_{i}(v_r^{({k})})$ should be satisfied for any prober and $g_{i}(\cdot)\geq \widehat{g}_i^*$ is bounded, there exists a natural number $k_0 < \infty$ such that $\vec{\xi}_i(k)$ always equals to $\vec{\xi}_{k_0}(k)$ =$\frac{\lambda_a}{k} (\chi \hbar_{ba},900\vartheta_k )$ when $k>k_0$. Let $v_{ba}$=$( \hbar_{ba},900 )$, $||\vec{\xi}_i(k)||$ satisfies the Robbins--Monro's conditions\cite{Newton2018Recent}. \emph{i.e.}, $\sum\limits_{k=1}^{k_0}||\vec{\xi}_i(k)||$ + $\sum\limits_{k=k_0}^{\infty} \frac{\lambda_a\mathbb{E}(|\chi|) ||v_{ba}||}{k}$ = $\infty$ and $\sum\limits_{k=1}^{k_0}||\vec{\xi}^2_i(k)|| + \sum\limits_{k=k_0}^{\infty}\frac{(\lambda_a \mathbb{E}(|\chi|) ||v_{ba}||)^2}{k^2}  < \infty$, where $\mathbb{E}(\cdot)$ denotes the expectation. This concludes the proof.
\end{myproof}

\vspace{0.1em}
In addition, since $||\vec{\xi}^{(k)}_{i}||$  in~\eqref{eqn_stepvec:big} is much larger than that in GD algorithm, the convergence speed may increase dramatically compared to the widely used  GD algorithm.

   \begin{figure}[t]
        \centering
        \includegraphics[width=3.1in, height=1.8in]{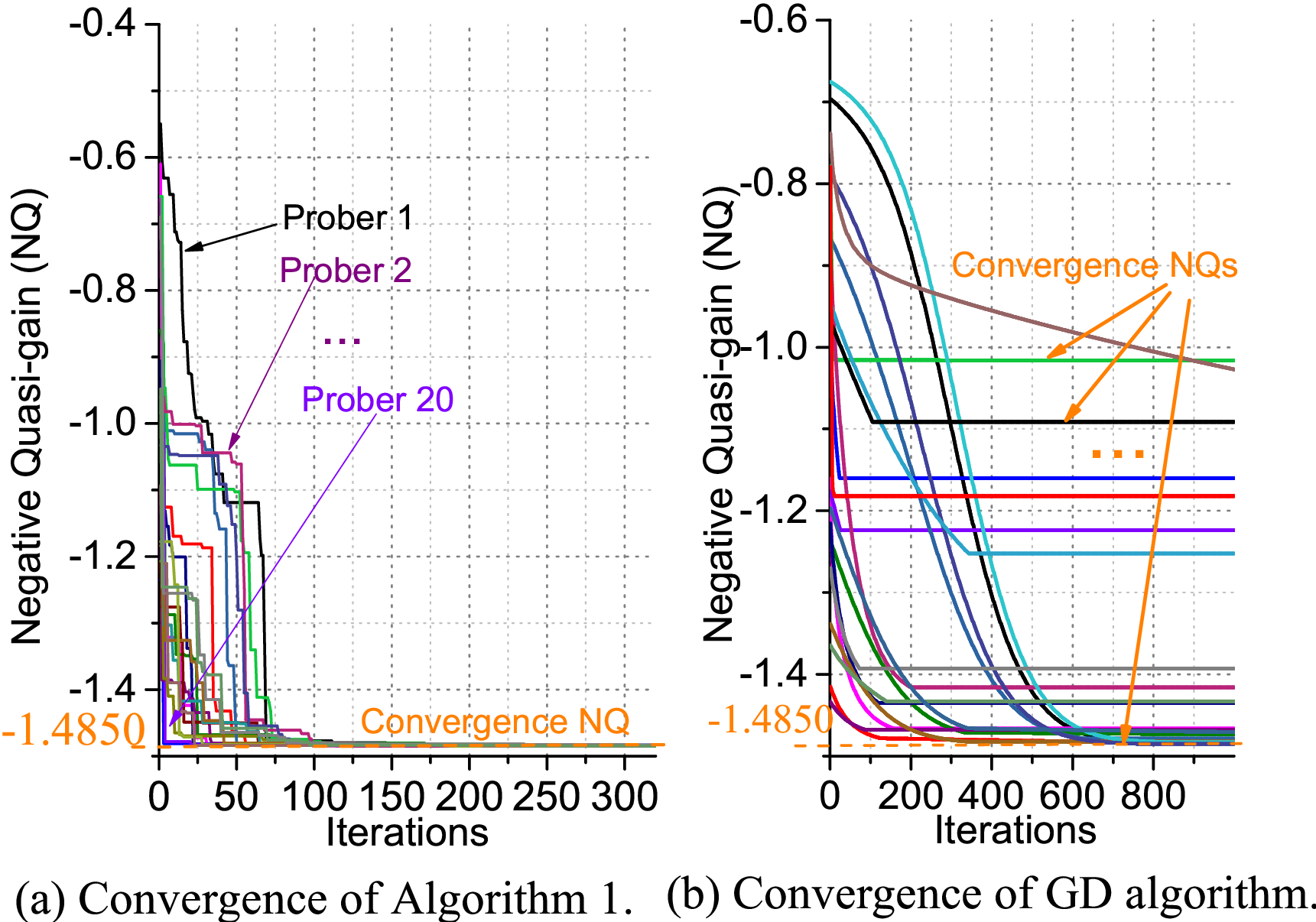}\\
        \vspace{-0.5em}
        \caption{Comparison of algorithm convergence under 20 random initial POs whose $x_r\in [x_{(a,i)}, x_{(b,i)}]$, $i$=$E_+$ and $\theta_r \in (-\infty^\circ, \infty^\circ)$. A color line indicates a searching track of a prober, $\lambda_a$=10, $\varepsilon_a$=0.1 and $\varpi$=3.
 }\label{fig_convergence}
    \vspace{-1.0em}
    \end{figure}

   \begin{figure}[t]
        \centering
        \includegraphics[width=3.42in, height=2.0in]{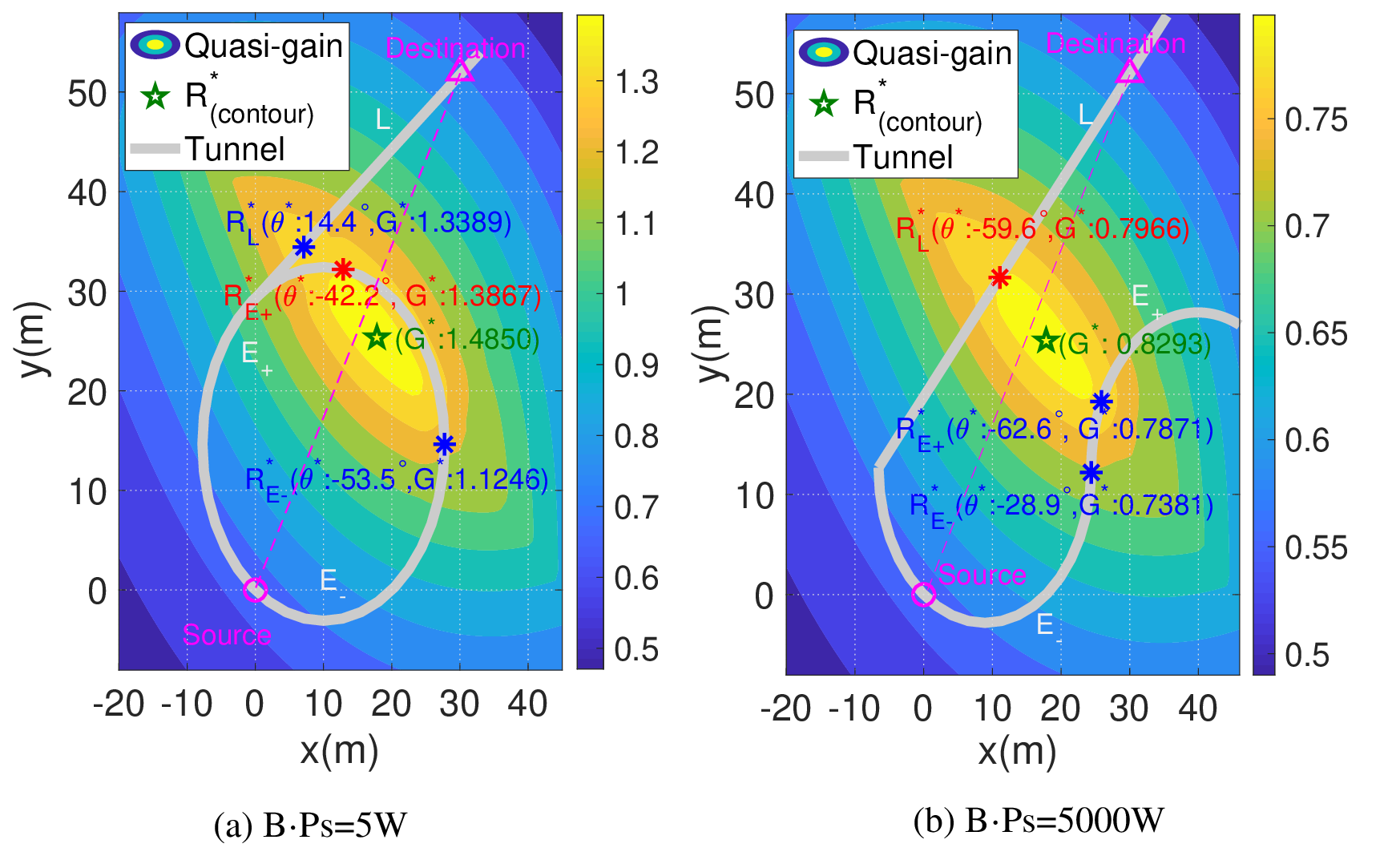}\\
        \vspace{-1.2em}
        \caption{The optimal antenna PO of the relay. R$_i^* $ is the optimal solution of the DF relay under  the sub-tunnel (equality) constraint $i \in \{E_+, E_-, L\}$. R$^*_{\rm (contour)}$ is the maximum (peak) of contour lines.
 }\label{fig_underPass}
      \vspace{-1.6em}
    \end{figure}

\section{Simulation}\label{sect sim}

 In this section, we verify our analysis by simulating the algorithm convergence and the tunnel constraint elimination. We assume the underground space is composed of dry soil whose conductivity  $\sigma$$=$$ 0.01S$$\cdot$ m$^{-1}$, permeability is $\mu$=4$\pi$$\times 10^{-7}$ H$\cdot$m$^{-1}$ \cite{kisseleff2013channel}. Similar to\cite{zhang2014cooperative}, the number of coil turns is $n$=$10$, coil radius is $a$=$0.5$m, the unit length resistance of the coil is $\rho_w$=$0.0166\Omega/$m, the load resistor $R_L$=$R_0$=$2\pi a n\rho_w $. We set the resonance frequency $f_0$=10 KHz, $\Gamma_p$$=$$0.5$, $\Gamma_{\rm fec}$$=$$0.5$, $\delta_{DF}$=$2\delta_{sd}$=6.4 and transmitting power $B$$\cdot$$P_r $=$B$$\cdot$$ P_s$.  The ambient noise level is assumed to $-103$ dBm\cite{Sun2010Magnetic}.

 To validate  Algorithm \ref{alg_apows},  we assume that the destination antenna is deployed with a specific  PO $\mathbf{v}_d$=$(30, 52, 30^\circ)$ which is 60m away from S.  Suppose S, D and the optimized relay should be deployed within the tunnel (as gray thick line in Fig.\ref{fig_road}) which is assumed to be divided into $E_+$ and $E_-$ with parameters ($x_c$=8.91m, $y_c$=12.67m, $R_a$=$R_b$=15.49m, $x_{a,e+}$=$x_{a,e-}$=$-6.58$m, $x_{b,e+}$=$x_{b,e-}$=24.4m) and $L$ with parameters ($x_{a,l}$=$-1$m, $x_{b,l}$$=$$33$m) by geometric approximation.
Fig.\ref{fig_convergence} shows the convergence of Algorithm \ref{alg_apows} and GD algorithm under the equality constraint $E_+$. We observe that all $\mathcal{M}$=$20$ probers  converge to the same point in Fig.\ref{fig_convergence}(a), while there exist multiple convergence points in Fig.\ref{fig_convergence}(b). These phenomena indicate that  Algorithm \ref{alg_apows} performs much better global search ability than widely used GD algorithm. Moreover, in Fig.\ref{fig_convergence}(b), most prober  lines drastically drop down to  the convergence point within 125 iterations since~\eqref{eqn_stepvec:big} is enabled. While in Fig.\ref{fig_convergence}(b), most probers drop to convergence point within 600 iterations due to the small step size. Thus, Algorithm \ref{alg_apows} also has a better convergence speed.

Fig.\ref{fig_underPass} shows two examples of solving the optimal antenna PO within other two tunnels through~\eqref{eqn_APOUS} and Algorithm \ref{alg_apows}. Using Algorithm \ref{alg_apows}, we first obtain the optimal antenna POs R$^*_{E+}$, R$^*_{E-}$ and R$^*_{L}$ under sub-tunnel constraints $E+$, $E-$ and $L$, respectively. Here R$^*_{E+}$, R$^*_{E-}$ and R$^*_{L}$ are within the sub-tunnels $E+$, $E-$ and $L$, respectively, which validates the tunnel constraint elimination.
Then, from R$^*_{E+}$, R$^*_{E-}$ and R$^*_{L}$, we choose the antenna PO (R$^*_{E+}$ in Fig.\ref{fig_underPass}(a)) with maximal cooperative quasi-gain larger than 1 ($G^* >1$) as the optimal PO of the DF relay for the antenna deployment which can improve the data rate performance. In a few cases (Fig.\ref{fig_underPass}(b)) with dramatically strong signals,  if we  find that the quasi-gain of the global optimal antenna PO (R$^*_{L}$ in Fig.\ref{fig_underPass}(b)) is smaller than 1, we use the direct MI communication.

\vspace{-0.4em}

\section{Conclusion}\label{sect conclusions}
In this paper, the potential of MI system with a DF relay within a tunnel is exploited by investigating the relay optimization problem with respect to antenna POs. This problem is subject to a tunnel constraint. We transform such constrained problem into unconstrained problems by geometric approximation. To solve the unconstrained problem, we propose a novel algorithm aiming at improving convergence and global search abilities.  Simulations indicate that the algorithm converges fast and is with excellent global search ability. Also, the optimization of the relay antenna can strikingly improve the  throughput of underground vehicle communications with weak signals.

\vspace{-0.1em}

\ifCLASSOPTIONcaptionsoff
  \newpage
\fi

\end{document}